# Production of Porous Glass-foam Materials from Photovoltaic Panel Waste Glass

Bui Khac Thach [1,2], Le Nhat Tan [1,2], Do Quang Minh [1,2,a)], Ly Cam Hung [3], Phan Dinh Tuan [3]

[1] Faculty of Materials Technology, Ho Chi Minh City University of Technology, 268 Ly Thuong Kiet St., Ward 14, District 10, Ho Chi Minh City 70000, Vietnam
[2] Vietnam National University Ho Chi Minh City, Linh Trung Ward, Thu Duc District, Ho Chi Minh City 70000, Vietnam
[3] Hochiminh City University of Natural Resources and Environment, 236B Le Van Sy St., Tan Binh District, Ho Chi Minh City 70000, Vietnam

[a)] Corresponding author: mnh_doquang@hcmut.edu.vn

**Abstract** The Solar energy production is growing quickly for the global demand of renewable one, decrease the dependence on fossil fuels. However, disposing of used photovoltaic (PV) panels will be a serious environmental challenge in the future decades since the solar panels would eventually become a source of hazardous waste. The potential of waste solar panel glass to generate porous glass material with the addition of $CaCO_3$ and water glass was assessed in this study. The porous glass firing temperature range, from 830°C - 910°C, was determined using a simulation of heating microscope technique. The created samples have the smallest volumetric density of 0.25 g/cm$^3$ and the largest water absorption of 303.08 wt.%. This indicates that the image analysis of samples during the heating process could be used to identify the firing temperature for better foaming, which was favorably indicated by specific physicochemical parameters. The created glass-foam materials with an apparent porosity up to 81.49% could be used as a water-retaining medium in hydroponic and aquaponic systems.

# 1 Introduction

Solar power generation is expanding rapidly and providing significant benefits. The estimated lifetime of photovoltaic (PV) modules is about 25 years. Therefore, in the coming decades, solar panels may eventually become a source of hazardous waste, and disposing of PV panels will be a crucial environmental issue [1]. Furthermore, most of the solid wastes from the used solar panels would be landfilled, coursing the contamination of soil and groundwater, damaging the ecosystem in the long term [2].

After being separated from PV modules, the glass from wasted solar panels is difficult to be recycled in floating or container glass furnaces due to its impurities. The procedure of purifying the glass from waste solar panels is complicated and expensive [1]. In thermal delamination, the ethylene vinyl acetate (EVA) is eliminated, and components including glass, aluminum frames, plastics and other components are separated [3]. Kang et al. [4] used organic solvents to recover glass from waste solar panels, after the panels were soaked in toluene for 2 days at 90°C, the tempered glass and PV cells are separated from the swollen and dissolved EVA. Pagnanelli et al. [5] recovered approximately 91% of the glass weight using two stages treatment: a physical treatment (triple crushing and thermal treatment) and a chemical treatment. Also, melting at temperature up to 1550°C in glass furnace consumes a large amount of energy and causes $CO_2$ and $SO_2$ emission [6].

There have been studies using waste glass flakes as a by-product in the production of ceramic materials, such as fired clay bricks and floor tiles [7, 8]. In addition, waste glass fragments have also been studied to make geopolymers [9], glass ceramics, etc. [10]. It has also been studied using solar panels as an aggregate for Portland cement [11]. Thanks to the pozzolanic activity of glass powder, a new material called "High-Performance Glass Concrete" has been created with the low water absorption, smooth surface, and high volumetric density. It also has rheological properties that enhance the workability of fresh concrete [12]. Waste glass and red mud have been mixed to make bricks by pressing the products and then sintering at the temperatures below 1000°C [13]. Waste glass mixed with $Ca(OH)_2$ in the ratio Ca/Si=0.83 can also form calcium silicate materials through hydrothermal treatment. As the hydrothermal time increased, the mechanical properties and density of the materials improved [14]. Wollastonite ($CaSiO_3$) material can be prepared at low temperature by using a mixture of rice husk ash, waste glass, and CaO by the hydrothermal process followed by calcination at 1000°C [15].

Porous glass is an application of vitrification typically accomplished by adding a gas-forming decomposing agent to the molten glass to create air bubbles into the glass matrix. Porous glass has desirable properties such as high porosity, low thermal conductivity, chemical resistance, water resistance, and non-flammability [16]. In general, foaming agents are divided into two categories: foaming via decomposing agents such as calcium carbonate ($CaCO_3$) [17], silicon carbide (SiC) [18], and gaseous combustions as carbon (C) etc. The volumetric density of glass foam tends to decrease as the initial particle size decreases, and rapidly decreases when the initial size of the glass powder reaches below 125 μm. The foaming is almost non-existent if the initial size of the glass powder is larger than 0.4 mm [19, 20]. Heating rates of 5 – 10°C/min are commonly used to avoid cracking due to too fast heating or premature gas generation due to too slow heating [21].

The properties of porous glass are fundamentally affected by the firing temperature, which is usually chosen between 700 and 900°C [22]. The porous glass has been demonstrated for its filtering ability to remove undesirable minerals from brackish water [23]. It must meet high strength, lightweight and chemical stability [24]. The material for insulation application can be fabricated with glass grain sizes of 75 - 150 μm and a sintering temperature of 800°C [25]. Besides, it creates an excellent hydroponic and aquaponic growing medium. After being irrigated, the pores help to absorb water quickly, hold water well, easily re-wet and have high porosity and aeration [26]. Furthermore, porous glass has many potential applications such as moisture control, electromagnetic absorption, lightweight bricks, and used as construction materials [27].

To make the porous glass, determining the firing temperature is very important. If the firing temperature is too high, the molten glass will seal the voids, but if the firing temperature is too low, the glass particles will not form stable bonds. Moreover, the calcination temperature needs to match the decomposition temperature of the foaming agents, so that the glass sample has high porosity while still having enough mechanical strength. Glass has no fixed melting point, but changes from solid to liquid through a soft temperature range. Experimentally, the softening temperature range was determined by Heating Microscopy (HM) [28]. The firing temperature for foaming glass must be chosen within the softening temperature range. Characteristically temperatures for the softening temperature range of glass are defined according to DIN 51730 (1998-4) / ISO 540 (1995-03-15) as follows:

 - Sintering temperature ($T_{sinter}$): the temperature at which the sample has had a dimensional shift of 5% in comparison to the initial image.

 - Softening temperature ($T_s$): the temperature for which round edges were visible (H = ¾ $H_o$).

 - Sphere temperature: the temperature for which the probe appeared like a sphere (H = D).

 - Hemisphere temperature: the temperature for which the height is half of the base (H = 1/2D).

 - Flow temperature ($T_f$): the temperature for which the sample is melted down to 1/3 of its initial height (H = 1/3$H_0$).

The softening temperature range ΔT is determined by $\Delta T = T_f - T_s$. The firing temperatures of the glass samples will be selected in this temperature range.

In this study, a simulation of the HM technique was performed. Characteristic temperatures in the glass softening range from waste solar panels with $CaCO_3$ foaming agent were determined by analyzing images taken with a digital camera. Then, the firing temperatures of the glass samples were selected in the glass softening range so that the $CaCO_3$ decomposed to pores, and the sintered glass samples had high enough mechanical strength.

## 2 Materials and methods

### *2.1 Raw Materials*

Broken PV panels were collected from Solar Vietnam JSC, Ho Chi Minh City, Vietnam. After using a heat gun to apply heat on the surface of the solar panels, the glass was manually separated from EVA layer. These separated materials contain waste glass fragments, pieces of solar cells and aluminum ribbons (Fig. 1a, b).

The waste glass was sorted and collected by hand, then ground using a ball mill in 6 hours. The powder then was sieved through 125 µm to remove EVA residues (Fig. 1c) and used to form the experimental samples. Commercially available $CaCO_3$ powders was used as high-temperature foaming agents and water glass was used as the initial binder. Their chemical compositions were analyzed using X ray fluorescence (XRF) with a loss on ignite at 950°C in 0.5 hour (ARL ADVANT'X, Thermo).

The particle size distributions of the ground waste glass and $CaCO_3$ was analyzed by Laser Diffraction Size Analysis (Mastersizer MS3000, Malvern Panalytical).

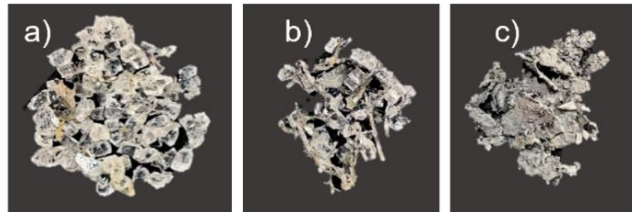

**Fig. 1** a) Waste glass fragments, b) Waste glass and aluminum ribbons, c) EVA residues

### *2.2 Experimental methods*

The Canon 700D digital camera with Lens 18-55 IS STM was used to take images of the specimen during the heating process with the heating rate of 10°C/min. The specimen with the mixture of 10% of $CaCO_3$ and 10% water glass in the proportion of the waste glass by weight was compressed into a cylindrical block with 3 mm high and 3 mm in diameter. The characteristic temperatures were selected by analyzing the behavior of green sample under heating process.

Raw materials were mixed in ratio of 2, 4, 6, 8 and 10% of $CaCO_3$ and 10% water glass, based on the weight of the waste glass. Tap water was used to control the humidity of mixture fixed at 20% (in wt.) then mixed for 5 minutes using an electric beater. After thorough mixing, the mixture was formed using a PVC mold about 12 mm in height and 18 mm in diameter. Then the samples were dried in the dryer (Venticell) at temperature of 110°C for 4 hours. Physical properties of the samples including volumetric density, water absorption and apparent porosity were determined following the standard of ASTM C830 (2011). The total porosity was obtained from the volumetric density and the sintered powder density with the following equations:

Total porosity = (1 – volumetric density / sintered powder density) x 100%

To ascertain the compressive strength of the glass foam, the cylindrical specimens of 15 mm high and 30 mm diameter were tested using a loading speed of 1.5 kN/min (DTU-900MHN, Daekyung Tech, Korea). The compressive strength test was illustrated in Fig 2.

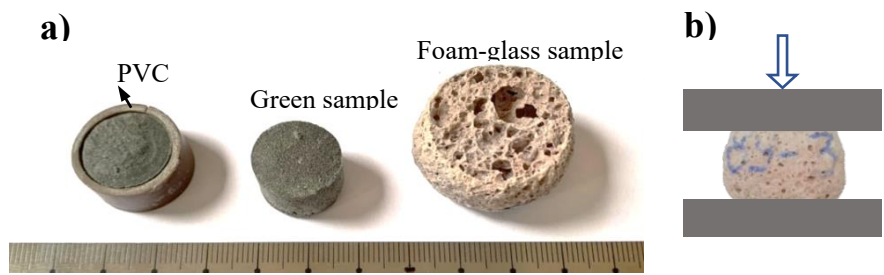

**Fig. 2** Images of a) PVC mold, green sample, and foam-glass sample, respectively, b) diagram of compressive strength test of foam-glass samples

## 3 Result and Discussions

### 3.1 Characteristics of raw materials

Chemical compositions of raw materials used to make the porous glass samples are shown in Table 1. The results of chemical analysis show that the glass from waste solar panels was based on soda-lime-silica glass system.

**Table 1.** The chemical composition (wt.%) of raw materials

| Material | $SiO_2$ | $Na_2O$ | CaO | $Al_2O_3$ | $Fe_2O_3$ | MgO | Others | LOI |
|---|---|---|---|---|---|---|---|---|
| Waste glass | 67.17 | 12.74 | 11.74 | 2.00 | 1.92 | 2.43 | 1.44 | 0.56 |
| $CaCO_3$ | - | - | 56.28 | - | 0.01 | 0.32 | 0.01 | 43.38 |
| Water glass | 69.15 | 27.82 | 0.40 | 1.66 | 0.07 | - | 0.90 | 0.00 |

The particle size distribution of the raw materials is shown as Fig. 3. According to the results of this analysis, the mean size ($d_{50}$) of the waste glass from solar panels and $CaCO_3$ powder is 35.0 μm and 18.1 μm, respectively.

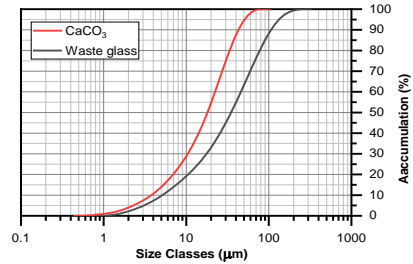

**Fig. 3** The particle size distributions of waste glass and $CaCO_3$

## 3.2 Investigation of foaming process

The thermal behaviors of sample were shown in Fig 4. The sintering temperature starts at 730°C. With increasing temperatures, the expansion of the sample took place at 770°C and maximum expansion 870°C thanks to the $CO_2$ release from the decomposition of $CaCO_3$. Then, the shape of the sample remained unchanged until 990°C - the softening point, and almost reached the sphere at 1000°C when the rounding of the edges is visible. As the temperature gets higher, the hemisphere and flow point could be observed at 1110°C and 1120°C, respectively.

Based on the behavioral characterization of the sample, the firing temperatures were chosen at 830, 850, 870, 890 and 910°C with the heating rate at 10°C/min for 15 minutes in an electrical kiln.

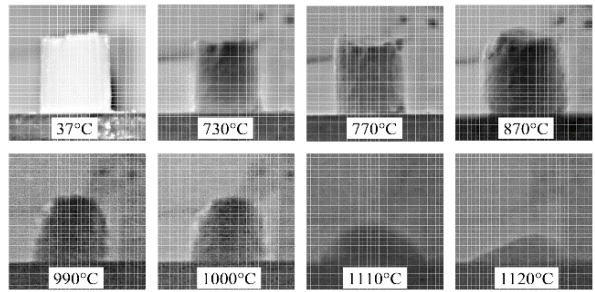

**Fig. 4** Images of the foaming process of foam-glass specimen.

## *3.3 Characteristic appearance of the foam samples*

The pore morphology of the two surfaces of the samples sintered at 870 °C for 15 min with different $CaCO_3$ contents is shown in Fig. 5. It can be seen that the increase of $CaCO_3$ makes the pore size of the sintered samples more evenly. Furthermore, the pore structure of the 10% weight $CaCO_3$ sample is considered to be uniform. The porosity distribution is not uniform because the bubbles rise to the top of the samples when the vitreous viscosity is low. A maze-like network of pores in glass is created when the intense gas released during their decomposition disrupts the walls of the individual pores [29]. The pore diameter was significantly reduced in samples where the $CaCO_3$ content increased by up to 10% by weight, similar results were reported by Souza et al [30] when eggshell was used as a foaming agent.

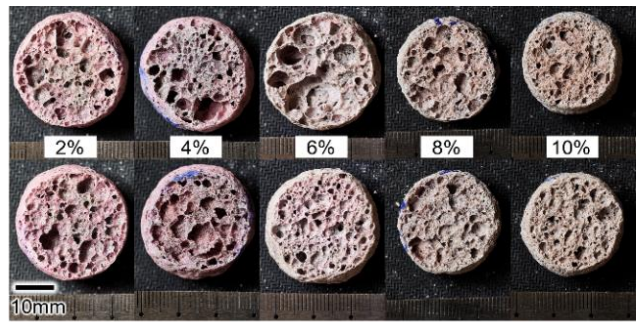

**Fig. 5** Images of the top (first row) and bottom (second row) of the samples heating at 870°C with different $CaCO_3$ contents from 2 to 10% (in wt.)

## *3.4 Physical and mechanical properties of sintered samples*

The physical properties of the test samples are shown in Fig. 6. The volumetric density of the sintered samples was increased with increasing $CaCO_3$ content from 2 to 10 wt.%. Samples with high total porosity have a low volumetric density and will therefore absorb a larger amount of water. Open porosity has a significant effect on water absorption. With the increase of open porosity, water absorption increases. According to the results, 870 °C was considered as the ideal sintering temperature to maximize the water absorption for all amounts of $CaCO_3$ corresponding to the maximum volume expansion.

Comparing the sintered samples with other porous glass produced from glass bottles and shells [22], the densities of the sintered samples were significantly lower.

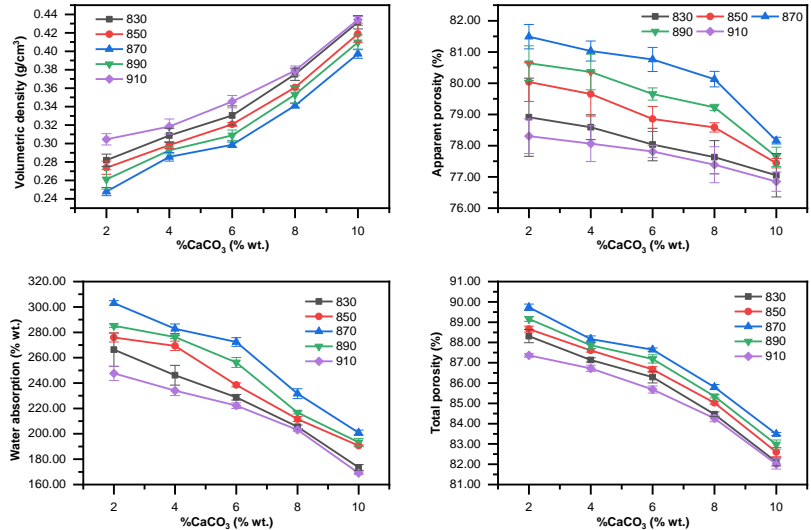

**Fig. 6** Influence of sintering temperature and CaCO$_3$ content on the physical properties of the glass foam samples.

At the low sintering temperature, 830°C, gas generation in the highly viscous glass matrix produces closed pores. In addition, incomplete sintering of the glass powder results in an almost porous mass. The partition walls between the pores are expected to deform at higher temperatures, 910°C, reducing the apparent porosity. As the CaCO$_3$ content increases, the volumetric density tends to increase. The results are similar to the reported study [31], especially the samples with 10% by weight CaCO$_3$.

The compressive strength of the samples with different amounts of CaCO$_3$ is shown in Fig. 7. It is clear that the 10% CaCO$_3$ samples had the highest compressive strength and the lowest porosity.

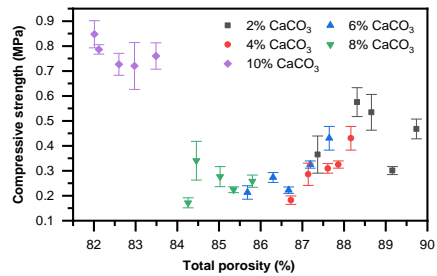

**Fig. 7** Total porosity versus compressive strength of samples with different CaCO$_3$ content (in wt.%)

## 4 Conclusion

Various samples of foam glass with different porosity have been formed from a mixture of $CaCO_3$, water glass, and glass powder from used solar panels. The method of prototyping with a 10 wt.% water glass binder and then casting it in a plastic mold has been shown suitable to create such types of materials.

The firing temperature for foaming was selected by image analysis of the sample during the heating process, favorably tailored to specific physicochemical properties.

Under the experimental conditions, the samples had the volumetric density of 0.25 - 0.43 g/cm$^3$ and water absorption of 160.02 - 303.08 wt.% as the heating temperature in the range of 830 - 910°C and the $CaCO_3$ contents of 2 – 10 wt.%.

The results indicated that as the $CaCO_3$ content increased, but the porosity decreased and the mechanical strength increased. Glass-foam-based materials with an apparent porosity of 76.85 to 81.49% can be used as a water-retaining material in hydroponic and aquaponic systems. It is also expected to be used as microbial media in water and wastewater treatments.

## Acknowledgements

The authors acknowledge financial support from the Ministry of Natural Resources and Environment of Vietnam through the Project coded 04/HĐ-BQLTDA-tm.
*Conflict of interest:* The authors declare no competing interests.

## References

1. Md. S. Chowdhury, K. S. Rahman, T. Chowdhury, N. Nuthammachot, K. Techato, Md. Akhtaruzzaman, S. K. Tiong, K. Sopian, and N. Amin, "An overview of solar photovoltaic panels' end-of-life material recycling," *Energy Strategy Rev.*, vol. 27, p. 100431, Jan. 2020, doi: 10.1016/j.esr.2019.100431.
2. J. I. Kwak, S.-H. Nam, L. Kim, and Y.-J. An, "Potential environmental risk of solar cells: Current knowledge and future challenges," *J. Hazard. Mater.*, vol. 392, p. 122297, Jun. 2020, doi: 10.1016/j.jhazmat.2020.122297.
3. E. Klugmann-Radziemska, P. Ostrowski, K. Drabczyk, P. Panek, and M. Szkodo, "Experimental validation of crystalline silicon solar cells recycling by thermal and chemical methods," *Sol. Energy Mater. Sol. Cells*, vol. 94, no. 12, pp. 2275–2282, Dec. 2010, doi: 10.1016/j.solmat.2010.07.025.
4. S. Kang, S. Yoo, J. Lee, B. Boo, and H. Ryu, "Experimental investigations for recycling

of silicon and glass from waste photovoltaic modules," *Renew. Energy*, vol. 47, pp. 152–159, Nov. 2012, doi: 10.1016/j.renene.2012.04.030.

5. F. Pagnanelli, E. Moscardini, G. Granata, T. Abo Atia, P. Altimari, T. Havlik, and L. Toro, "Physical and chemical treatment of end of life panels: An integrated automatic approach viable for different photovoltaic technologies," *Waste Manag.*, vol. 59, pp. 422–431, Jan. 2017, doi: 10.1016/j.wasman.2016.11.011.
6. P. T. Kien, K. D. T. Kien, and D. Q. Minh, "Research on wasted glass as non-firing brick using hydrothermal method," *J. Sci. Technol.*, vol. 52, no. 4A, pp. 198-204., Sep. 2014.
7. K.-L. Lin, T.-C. Lee, and C.-L. Hwang, "Effects of sintering temperature on the characteristics of solar panel waste glass in the production of ceramic tiles," *J. Mater. Cycles Waste Manag.*, vol. 17, no. 1, pp. 194–200, Mar. 2014, doi: 10.1007/s10163-014-0240-3.
8. J. Jimenez-Millan, I. Abad, R. Jimenez-Espinosa, and A. Yebra-Rodriguez, "Assessment of solar panel waste glass in the manufacture of sepiolite based clay bricks," *Mater. Lett.*, vol. 218, pp. 346–348, May 2018, doi: 10.1016/j.matlet.2018.02.049.
9. H. Hao, K.-L. Lin, D. Wang, S.-J. Chao, H.-S. Shiu, T.-W. Cheng, and C.-L. Hwang, "Elucidating characteristics of geopolymer with solar panel waste glass," *Environ. Eng. Manag. J.*, vol. 14, no. 1, pp. 79–87, 2015, doi: 10.30638/eemj.2015.010.
10. V. Savvilotidou, A. Kritikaki, A. Stratakis, K. Komnitsas, and E. Gidarakos, "Energy efficient production of glass-ceramics using photovoltaic (P/V) glass and lignite fly ash," *Waste Manag.*, vol. 90, pp. 46–58, May 2019, doi: 10.1016/j.wasman.2019.04.022.
11. K. Máčalová, V. Václavík, T. Dvorský, R. Figmig, J. Charvát, and M. Lupták, "The Use of Glass from Photovoltaic Panels at the End of Their Life Cycle in Cement Composites," *Materials*, vol. 14, no. 21, p. 6655, Nov. 2021, doi: 10.3390/ma14216655.
12. J. Esmaeili and A. Oudah AL-Mwanes, "A review: Properties of eco-friendly ultra-high-performance concrete incorporated with waste glass as a partial replacement for cement," *Mater. Today Proc.*, vol. 42, pp. 1958–1965, 2021, doi: 10.1016/j.matpr.2020.12.242.
13. D. Q. Minh, N. H. Thang, and P. T. Kien, "Research on manufacturing decorative tiles from cullet and red mud," in *Green Material and Green Technology for Green "MONOZUKURI,"* Hanoi University of Science and Technology, Hanoi, Vietnam, Nov. 2013, pp. 75–81.
14. P. T. Kien, N. T. H. Phong, N. H. Thang, K. D. T. Kien, D. Q. Minh, L. H. Anh, and P. T. T. Trinh, "Wasted cullet in glass technology used as ecomaterials," in *Collaboration and Exchange in Advanced Science and Technology*, Can Tho University, Vietnam, Sep. 2014, pp. 73–78.
15. P. Trung Kien, T. Thi Thien Ly, P. Thi Lan Thanh, T. Pham Quang Nguyen, N. Hoc Thang, and M. M. A. B. Abdullah, "A Novel Study on Using Vietnam Rice Hush Ash and Cullet as Environmental Materials," *MATEC Web Conf.*, vol. 97, p. 01118, 2017, doi: 10.1051/matecconf/20179701118.
16. G. Scarinci, G. Brusatin, and E. Bernardo, "Glass Foams," in *Cellular Ceramics (Structure, Manufacturing, Properties and Applications)*, M. Scheffler and P. Colombo, Eds. Weinheim, FRG: Wiley-VCH Verlag GmbH & Co. KGaA, 2006, pp. 158–176. doi:

10.1002/3527606696.ch2g.

17. M. Zhu, R. Ji, Z. Li, H. Wang, L. Liu, and Z. Zhang, "Preparation of glass ceramic foams for thermal insulation applications from coal fly ash and waste glass," *Constr. Build. Mater.*, vol. 112, pp. 398–405, Jun. 2016, doi: 10.1016/j.conbuildmat.2016.02.183.
18. J. Bai, X. Yang, S. Xu, W. Jing, and J. Yang, "Preparation of foam glass from waste glass and fly ash," *Mater. Lett.*, vol. 136, pp. 52–54, Dec. 2014, doi: 10.1016/j.matlet.2014.07.028.
19. V. Ducman and M. Kovačević, "The Foaming of Waste Glass," *Key Eng. Mater.*, vol. 132–136, pp. 2264–2267, Apr. 1997, doi: 10.4028/www.scientific.net/KEM.132-136.2264.
20. J. König, R. R. Petersen, and Y. Yue, "Influence of the glass particle size on the foaming process and physical characteristics of foam glasses," *J. Non-Cryst. Solids*, vol. 447, pp. 190–197, Sep. 2016, doi: 10.1016/j.jnoncrysol.2016.05.021.
21. G. Scarinci, G. Brusatin, and E. Bernardo, "Glass Foams," in *Cellular Ceramics (Structure, Manufacturing, Properties and Applications)*, M. Scheffler and P. Colombo, Eds. Weinheim, FRG: Wiley-VCH Verlag GmbH & Co. KGaA, 2006, pp. 158–176. doi: 10.1002/3527606696.ch2g.
22. N. A. N. Hisham, M. H. M. Zaid, S. H. A. Aziz, and F. D. Muhammad, "Comparison of Foam Glass-Ceramics with Different Composition Derived from Ark Clamshell (ACS) and Soda Lime Silica (SLS) Glass Bottles Sintered at Various Temperatures," *Materials*, vol. 14, no. 3, p. 570, Jan. 2021, doi: 10.3390/ma14030570.
23. Sulhadi, Susanto, A. Priyanto, A. Fuadah, and M. P. Aji, "Performance of Porous Composite from Waste Glass on Salt Purification Process," *Procedia Eng.*, vol. 170, pp. 41–46, 2017, doi: 10.1016/j.proeng.2017.03.008.
24. X. Fang, Q. Li, T. Yang, Z. Li, and Y. Zhu, "Preparation and characterization of glass foams for artificial floating island from waste glass and Li2CO3," *Constr. Build. Mater.*, vol. 134, pp. 358–363, Mar. 2017, doi: 10.1016/j.conbuildmat.2016.12.048.
25. S. S. Owoeye, G. O. Matthew, F. O. Ovienmhanda, and S. O. Tunmilayo, "Preparation and characterization of foam glass from waste container glasses and water glass for application in thermal insulations," *Ceram. Int.*, vol. 46, no. 8, pp. 11770–11775, Jun. 2020, doi: 10.1016/j.ceramint.2020.01.211.
26. M. Raviv, J. H. Lieth, and A. Bar-Tal, Eds., *Soilless culture: theory and practice*, Second edition. Amsterdam: Academic Press, an imprint of Elsevier, 2019.
27. M. Flood, L. Fennessy, S. Lockrey, A. Avendano, J. Glover, E. Kandare, and T. Bhat, "Glass Fines: A review of cleaning and up-cycling possibilities," *J. Clean. Prod.*, vol. 267, p. 121875, Sep. 2020, doi: 10.1016/j.jclepro.2020.121875.
28. C. Venturelli, "Heating Microscopy and its Applications," *Microsc. Today*, vol. 19, no. 1, pp. 20–25, Jan. 2011, doi: 10.1017/S1551929510001185.
29. Yu. A. Spiridonov and L. A. Orlova, "Problems of Foam Glass Production," *Glass Ceram.*, vol. 60, no. 9/10, pp. 313–314, Sep. 2003, doi: 10.1023/B:GLAC.0000008234.79970.2c.
30. M. T. Souza, B. G. O. Maia, L. B. Teixeira, K. G. de Oliveira, A. H. B. Teixeira, and A. P. Novaes de Oliveira, "Glass foams produced from glass bottles and eggshell wastes," *Process Saf. Environ. Prot.*, vol. 111, pp. 60–64, Oct. 2017, doi:

10.1016/j.psep.2017.06.011.

31. J. König, R. R. Petersen, and Y. Yue, "Influence of the glass–calcium carbonate mixture's characteristics on the foaming process and the properties of the foam glass," *J. Eur. Ceram. Soc.*, vol. 34, no. 6, pp. 1591–1598, Jun. 2014, doi: 10.1016/j.jeurceramsoc.2013.12.020.